\documentclass[prd,twocolumn,showpacs,superscriptaddress,floatfix]{revtex4}

\usepackage{graphicx,psfrag,amsmath,amssymb,amsfonts,bbm,latexsym,color,dcolumn,bm}

\begin{document}

\title{Higgs-induced spectroscopic shifts near strong gravity sources}

\author{Roberto Onofrio}
\email{onofrior@gmail.com}
\affiliation{Dipartimento di Fisica ``Galileo Galilei'', 
Universit\`a  di Padova, Via Marzolo 8, Padova 35131, Italy}

\affiliation{ITAMP, Harvard-Smithsonian Center for Astrophysics, 
60 Garden Street, Cambridge, MA 02138, USA}
\date{\today}

\begin{abstract}
We explore the consequences of the mass generation due to 
the Higgs field in strong gravity astrophysical environments. 
The vacuum expectation value of the Higgs field is predicted 
to depend on the curvature of spacetime, potentially giving 
rise to peculiar spectroscopic shifts, named hereafter ``Higgs shifts.'' 
Higgs shifts could be searched through dedicated multiwavelength 
and multispecies surveys with high spatial and spectral resolution 
near strong gravity sources such as Sagittarius A$^*$ or 
broad searches for signals due to primordial black holes. 
The possible absence of Higgs shifts in these surveys 
should provide limits to the coupling between the Higgs 
particle and the curvature of spacetime, a topic of interest 
for a recently proposed Higgs-driven inflationary model. 
We discuss some conceptual issues regarding the coexistence  
between the Higgs mechanism and gravity, especially for their 
different handling of fundamental and composite particles.

\end{abstract}


\maketitle

\section{Introduction}

One of the most important predictions of the standard model of
particle physics is the existence of a scalar particle, called 
the Higgs particle, responsible for the spontaneous symmetry breaking 
of the electroweak sector, providing a dynamical mechanism to generate 
the mass of the intermediate vector bosons $W^{\pm}$ and $Z^0$ and of 
all fundamental fermionic matter fields \cite{Higgs}. 
The identification of the Higgs particle is considered an important 
milestone for the final validation of the standard model, and is the 
primary focus of research planned at the highest energy accelerators 
such as the Large Hadron Collider at CERN. 

Although several implications of the Higgs coupling to fermions have 
been discussed in detail, little attention has been devoted so far,
to our knowledge, to the fact that the Higgs particle should also play 
a crucial role in gravitational phenomena, provided that it satisfies 
the equivalence principle. If the Higgs field is coupled to the
spacetime metric, its vacuum expectation value should differ from 
the one in a flat spacetime. As discussed in Sec. II (see also 
\cite{Onofrio} for a preliminary account), different values for 
the mass of particles such as electrons and protons should then 
occur in the same region, with consequences for the energy levels 
of bound states of spectroscopic relevance. 
Peculiar ``Higgs shifts'' in the emission or absorption spectrum of atoms 
are expected, and it should be possible to distinguish them amidst 
the usual Doppler, gravitational, and cosmological shifts using  
multispecies spectroscopic analysis. 
In Sec. III we discuss in more detail what we consider the 
most promising cases for the observation of Higgs shifts 
from supermassive black holes near the Galactic center or 
primordial black holes $-$ in particular, molecular tracers 
and neutral hydrogen in interstellar clouds and spectra of a star 
with high eccentricity, x-ray and $\gamma$-ray narrow lines. 
In the conclusions, we comment on general features of the 
Higgs-curvature connection, in particular potential instabilities 
due to the metric backreaction and the difference between Higgs 
physics and general relativity in dealing with the elementary or 
composite nature of particles.

\section{Higgs field in curved spacetime}

Quantum field theory in curved spacetime has been studied for several 
decades for both noninteracting and interacting fields (see
\cite{Birrell} for an overview). 
The Lagrangian density for an interacting scalar field in a 
generic curved spacetime $g^{\mu \nu}$ is written as:
\begin{equation}
{\cal L}= \sqrt{-g} \left[\frac{1}{2} g^{\mu \nu} 
\partial_\mu \phi \partial_\nu \phi - \frac{1}{2} (\mu^2 +\xi R) \phi^2 
-\frac{\lambda}{4}\phi^4\right],
\end{equation}
where $\mu$ and $\lambda$ are the mass parameter and the self-coupling quartic 
coefficient of the scalar field, respectively, $g$ is the determinant of 
the metric $g^{\mu\nu}$, and $\xi$ is a coefficient representing the coupling 
strength between the scalar field $\phi$ and the Ricci scalar $R$.
This last coefficient is considered as a free parameter in all models analyzed 
so far, and only two prescriptions have been suggested on theoretical grounds.
The so-called {\sl minimal} coupling scenario simply assumes $\xi=0$.  
This however is unnatural if the scalar field represents a Higgs field $-$ leaving 
aside the doublet nature of the latter due to the $SU(2)_L\otimes U(1)_Y$ gauge 
symmetry which will be irrelevant in the following discussion. 
Indeed, if we believe that the standard model at some energy will 
merge with gravitation, we expect an interaction term between 
metric invariants and the scalar field. A minimal coupling instead 
minimizes the crosstalk between the standard model and the gravitational 
sectors, as in this case they will be only related via the metric tensor-scalar 
field kinetic term in Eq. (1). 
Moreover, this choice is not stable against quantum corrections \cite{Linde}, 
as confirmed by studying renormalization group features \cite{Birrell}.
Proper renormalization behavior is instead fulfilled for a {\sl conformal} 
coupling of $\xi=1/6$, which has been shown to be a fixed point of the 
renormalization group equations \cite{Birrell,Freedman,Salopek}. 
It would be highly desirable to extract the Higgs-curvature coupling 
coefficient $-$ or to obtain at least upper bounds $-$ from the phenomenological 
analysis of particle observables in the presence of strong gravity, and 
from now on we discuss a possible scenario towards this direction.
  
In the spontaneously broken phase the Higgs field develops a vacuum 
expectation value $v_0={(-\mu^2/\lambda)}^{1/2}$ in flat spacetime, 
with the masses of the elementary particles directly proportional 
to $v_0$ via the Yukawa coefficients of the fermion-Higgs 
Lagrangian density term, $m_i=y_i v_0/\sqrt{2}$. 
In a curved spacetime instead, the effective coefficient 
of the Higgs field $\mu^2 \mapsto \mu^2+\xi R$, and the vacuum 
expectation value of the Higgs field will become spacetime 
dependent through the curvature scalar as:
\begin{equation}
v=\sqrt{-\frac{\mu^2+\xi R}{\lambda}} \simeq v_0 \left(1+\frac{\xi R}{2 \mu^2}\right),
\end{equation}
where the last expression holds in a weak-curvature limit. 

In the case of an elementary particle, such as the electron, provided 
that the Yukawa couplings $y_i$ are constants yet to be determined 
$-$ presumably from algebraic or group-theoretic arguments of an underlying 
fundamental theory embedding the standard model $-$ the mass $m_i$ will 
be simply changed proportionally to the Higgs vacuum expectation value, so that 
\begin{equation}
\delta m_i=\frac{y_i}{\sqrt{2}} (v-v_0)\simeq \frac{y_i \xi R
  v_0}{2^{3/2}\mu^2} = \frac{\xi R}{2 \mu^2} m_i.  
\end{equation}
The situation for composite particles such as protons and neutrons is
more involved. We assume that their masses are made of a flavor-dependent 
contribution proportional to the masses of the three valence quarks
determined by the Higgs coupling, and a color-symmetrical term only 
dependent on the quark-quark and quark-gluon interaction, {\it i.e.} 
proportional to the QCD constant $\Lambda_{QCD} \simeq 300$ MeV. 
The latter term dominates for lighter, relativistic quarks constituting 
the {\sl valence} component of protons and neutrons. 
Then, due to the universality of the QCD coupling constant for 
different flavors and for all gluons exchange, we can parametrize 
the proton and neutron masses in terms of flavor-dependent and 
flavor-independent parts as:
\begin{eqnarray}
m_p & = &  (2 y_u+y_d) v/\sqrt{2}+m_{QCD}, \nonumber \\ 
m_n & = & (y_u+2y_d) v/\sqrt{2}+m_{QCD},
\end{eqnarray}
where $y_u$ and $y_d$ are the Yukawa couplings of the up and down quarks, and 
$m_{QCD}$ is a flavor-independent contribution related to the gluon binding 
energy, depending on $\Lambda_{QCD}$.
For a generic atom of atomic number $Z$ and atomic mass $A$ we then obtain:
\begin{eqnarray}
& & M(A,Z) = Z m_p+(A-Z)m_n= \nonumber \\\
& & \frac{1}{\sqrt{2}}[y_u (Z+A)+y_d(2A-Z)]v+Am_{QCD},
\end{eqnarray}
where we have neglected to first approximation the contributions of the electron 
mass, the electron-nucleus binding energy, and the nucleon-nucleon binding energy. 
The purely QCD-dependent mass term should be independent on the curvature 
of spacetime, since otherwise the gluon could acquire a mass giving rise 
to the explicit breaking of the color symmetry. 
This is analogous to the case of the other unbroken symmetry of 
the standard model, $U(1)_\mathrm{em}$, which leads to the electric charge 
conservation even in a generic curved spacetime. 
By considering the Yukawa couplings $y_u$ and $y_d$ as determining the current 
quark masses $m_u$ and $m_d$ (with central values quoted in the Particle Data 
Group of 2.25 and 5 MeV, respectively), it is evident that for composite 
states of quarks such as protons and neutrons and their combinations, the 
flavor/Yukawa coupling independent term dominates, and the effect of curved 
spacetime is therefore strongly suppressed.
Therefore, the possibility of detecting Higgs shifts in atomic and molecular
spectroscopy relies on the fact that electronic transitions depend primarily 
on the mass of the electron, while molecular transitions due to vibrational 
or rotational degrees of freedom depend upon the mass of the nuclei. 
While the electron mass is directly proportional to the Yukawa 
coupling to the Higgs particle, the mass of the nuclei is mainly 
due to the contribution of its proton and neutron constituents, which  
in turn depends mainly on the color binding energy. 
We therefore expect that molecular transitions will not be affected 
by the Higgs shifts to leading order, unlike electronic transitions. 

In the relevant example of atomic hydrogen spectroscopy, the spectral 
lines depend on the reduced mass $\mu_H=m_e m_p/(m_e+m_p)$
and ultimately, due to the large mass ratio $m_p/m_e$, on the electron mass. 
At the molecular level, unless electronic transitions are excited, the 
Higgs shift is shown to be negligible even in the most favorable 
case of pyramidal molecules such as ammonia, for which tunneling 
provides exponentially higher sensitivity to the change in masses of the atoms. 
In particular, in the case of the nitrogen atom constituting 
the ammonia molecule, we have:
\begin{equation}
M_\mathrm{N}=\frac{21}{\sqrt{2}}(y_u+y_d)v+14 m_{QCD},
\end{equation}
with the effective mass for the inversion spectrum of ammonia equal to 
$\mu_\mathrm{NH_3}=3M_\mathrm{H} M_\mathrm{N}/(3M_\mathrm{H}+M_\mathrm{N})$. 
For Yukawa couplings of $y_e=2.89 \times 10^{-6}$, $y_u=1.27 \times 10^{-5}$,
$y_d=2.83 \times 10^{-5}$, and a pure gluonic contribution of $m_{QCD}=928$ MeV, 
mass shifts of $\delta\mu_\mathrm{H}/\mu_\mathrm{H}= 4 \times 10^{-3}$ for hydrogen 
and $\delta \mu_\mathrm{NH_3}/\mu_\mathrm{NH_3}= 3.4 \times 10^{-5}$ for ammonia 
are obtained for a variation of $\delta v= 1$ GeV around $v_0$= 250 GeV 
($\delta v/v_0=4 \times 10^{-3}$). 
Therefore, it is clear that, even if the ammonia inversion spectrum is 
in principle more sensitive (by a factor $\simeq 4 \div 5$ as discussed 
in \cite{Flambaum}) to the masses of its constituents than spectra 
from other molecular and nonpyramidal species, under the 
hypothesis that $m_{QCD}$ does not couple to the Higgs vacuum its 
sensitivity does not match the one of atomic hydrogen.

\section{Astrophysical considerations}

We now discuss qualitatively the possibility of observing Higgs shifts
from astrophysical objects. This implies a number of restrictive 
hypotheses both on the gravitational sources and their coupling 
to the Higgs particle, and on the detectability of the Higgs 
shift amidst other sources of wavelength shift. 
As remarked above, it is important to detect both spectroscopic 
lines due to electronic transitions and nuclear (vibrational or 
rotational) transitions. This is difficult to achieve in the 
same region of space from the same species for a gas at thermal 
equilibrium, due to the large energy scale difference required 
for effectively producing these excitations. A comparative analysis 
of wavelength shifts from different species seems then necessary.  
This should allow for discrimination from the Doppler shift and 
the purely gravitational shift. The Doppler shift should be the 
same for molecules belonging to the same comoving cloud, while 
the wavelength shift expected from general relativity will act 
universally on all particles, so unlike the Higgs shift it will 
not distinguish between fundamental particles and interactions 
binding energies.

A further difficulty is that the Ricci scalar $R$ is zero in the 
case of symmetrical gravitational sources, which are described by 
the Schwarzschild or the Kerr metric. We will then make the hypothesis 
that the Higgs field couples to another scalar invariant, for instance 
the Kretschmann invariant defined as 
$K_1=R_{\mu\nu\rho\sigma}R^{\mu\nu\rho\sigma}$, where
$R^{\mu\nu\rho\sigma}$ is the Riemann curvature tensor. 
This invariant plays an important role in quadratic theories of gravity
\cite{Deser,Stelle,Hehl} and more in general in modified $f(R)$ theories 
\cite{Sotiriou}. In the case of the Schwarzschild metric the 
Kretschmann invariant is $K_1=12 R_s^2/r^6$, with $R_s$ the Schwarzschild 
radius $R_s=2 GM/c^2$, and $r$ the distance from the center of the mass $M$. 
If we replace the Higgs-Ricci curvature coupling term $\xi\phi^2 R/2$ in 
Eq. (1) with a  Higgs-Kretschmann coupling term we obtain the modified 
Lagrangian density
\begin{equation}
{\cal L}_K= \sqrt{-g} \left[\frac{1}{2} g^{\mu \nu} 
\partial_\mu \phi \partial_\nu \phi - \frac{1}{2} (\mu^2 +\xi^\prime K_1^{1/2}) \phi^2 
-\frac{\lambda}{4}\phi^4\right],
\end{equation}
in which the curvature-scalar interaction term appears proportional to 
$K_1^{1/2}$ for dimensional reasons. This could appear problematic in 
regions of weak spacetime curvature where $K_1 \rightarrow 0$ since 
divergencies may occur, but in the context of spacetime regions analyzed here 
this coupling may be considered as arising from an effective interaction 
Lagrangian valid for strong and static gravitational fields.
In this case the mass term $\mu^2$ maps onto 
\begin{equation}
\mu^2 \mapsto \mu^2(1+2\sqrt{3}\xi^\prime R_s \lambda_\mu^2/r^3),
\end{equation}
where, in preparation for concrete estimates, we have introduced the
Compton wavelength associated to the Higgs mass parameter $\mu$ as
$\lambda_\mu=\hbar/(\mu c)$. Assuming a Higgs mass of 160 GeV and 
a vacuum expectation value of $v_0$=250 GeV, we obtain a Compton
wavelength for the Higgs mass parameter $\lambda_\mu \simeq 2 \times 
10^{-18}$ m: this is the length scale with which the Kretschmann 
invariant has to be confronted in any astrophysical setting. 

If we imagine collecting electromagnetic signals emitted from the innermost 
stable orbit of a Schwarzschild black hole, assuming that the Kretschmann invariant 
does not perturb significantly the stability analysis of black holes, we obtain for 
$r=3 R_s$, $K_1^{1/2}=(4/243)^{1/2} R_s^{-2}$. The frequency shift is therefore 
inversely proportional to the square of the Schwarzschild radius, and it gets 
larger by considering rotating black holes due to the smaller innermost stable 
orbits allowed in the Kerr metric \cite{Henry}. For supermassive black holes 
such as the one located in our Galaxy, Sagittarius A$^*$ with an estimated mass of 
$M \simeq 2.6 \times 10^6$ solar masses \cite{Morris,Eckart,Genzel,Ghez, Reid}, 
the Schwarzschild radius is equal to $R_s=\simeq 8 \times 10^9$ m. 
For a solar mass black hole we obtain $K^{1/2}=1.5 \times 10^{-8}$ m$^{-2}$. 
In the two cases the product $\lambda_{\mu}^2 K_1^{1/2}$ is 
$\simeq 8 \times 10^{-57}$ and $\simeq 6 \times 10^{-44}$ respectively, leading 
to tiny Higgs shifts quite far from what can be achieved with any foreseeable 
survey unless quite large values of the Higgs-curvature coupling parameter 
$\xi^\prime$ are allowed. If we consider mini black holes with a mass of the 
order of $10^{11}$ kg, which should survive evaporation via quantum tunneling 
\cite{Hawking,Carr1,Carr2}, we obtain $R_s \simeq 10^{-16}$ m, and 
$\delta m_i/m_i \simeq 2 \times 10^{-5} \xi^{\prime}  \simeq \delta \nu/\nu$. 
Recent surveys of molecular clouds, for instance containing ammonia 
\cite{Henkel,Wilson}, have a spectral sensitivity corresponding to 
a Doppler shift of about 2-3 km/s, {\it i.e.} $\delta \nu/\nu \simeq
10^{-5}$, comparable to the expected estimates based on mini black holes. 
If the same spectral sensitivity could be 
maintained in a broad survey of other spectroscopic transitions, upper 
limits of the order of $\xi^{\prime} \simeq 1$ could be achieved.

In spite of the pessimistic estimates reported above, it may be worthwhile 
to perform surveys near the Galactic center, especially keeping in 
mind the absolute lack of information on the Higgs-Kretschmann coupling 
$\xi^{\prime}$. With a 1 pc resolution survey one should be able to 
obtain spectra of atoms or molecules at a distance of $r \simeq 2 
\times 10^{16}$ m from the Galactic center. While detailed surveys 
of the Galactic center have been performed for various molecular 
species such as for instance NH$_3$ \cite{Kaifu,Gusten,Ho,McGary1,McGary2}, 
CO \cite{Liszt}, H$_2$CO \cite{Bieging,Gusten1}, and multispecies
\cite{Armstrong,Coil,Wright}, observation of atomic lines 
from the same region is difficult due to the strong absorption at 
optical wavelengths. This issue may be circumvented by focusing on 
high-precision observations of the 21 cm line of neutral hydrogen
which still depends on the electron-to-proton mass ratio. A further 
refinement on this proposal is obtained by monitoring neutral 
hydrogen surrounding stars with highly eccentric orbits around 
Sagittarius A$^*$. This should provide clearer signatures, especially 
in regard to a possible temporal variability of the 21 cm line 
related to the proximity of the star to the source of strong gravity. 

The presence of spectroscopic shifts related to the Higgs field 
could also be investigated in high-energy astrophysics phenomena. 
For instance, there should be a further contribution in the redshift
of the K$_\alpha$ emission line from ionized iron of stars orbiting 
in proximity of the source of spacetime curvature \cite{Tanaka}.
Another possibility is the detection of shifts in the annihilation 
spectrum near the Galactic center. Recent surveys have been performed 
with an energy resolution $\Delta E/E=1.47 \times 10^{-4}$ at the 
positron annihilation peak \cite{Churazov}. In this case it is 
crucial to achieve a high angular resolution of the detector, since 
the putative shifted signal from the Galactic center will be otherwise
smeared out by the nearby unshifted contributions. The intrinsic 
resolution of the 511 keV peak is limited by the environmental 
temperature around the Galactic center, estimated to be 
$T \leq 5 \times 10^4$ K \cite{Bussard}, which leads to a relative energy 
spread at the annihilation peak of $K_B T/E_{\gamma} \simeq 10^{-5}$.
With a measured positronium fraction close to unity (0.93 $\pm$ 0.04
from \cite{Kinzer}) and in the presence of neutral H or H${}_2$ gases, 
the influence of external magnetic or electric fields on the
annihilation spectrum should be minimized. A comparative analysis 
between signals for electron-positron and proton-antiproton annihilation
from strong gravity sources using telescopes with both large energy
and angular resolution (with Fermi/GLAST being the best candidate 
available now for the hadronic annihilation channel), might allow a 
detailed test of the presence of Higgs shifts.  

\section{Conclusions}

We have discussed the interplay between the Higgs particle and the 
curvature of spacetime and the possibility of observing peculiar
spectroscopic shifts from strong gravity astrophysical sources. 
Some final comments are in order. Although the discussion relies on 
strong gravity being associated to a nonzero Ricci scalar, or
coupling through the Kretschmann invariant, the main message 
discussed in this note is to search for frequency shifts which
discriminate between transitions associated to electronic or 
baryonic states. While we have focused on sources of astrophysical 
interest, similar considerations could be extended in a cosmological 
framework, for instance by looking at the presence of specific frequency 
shifts in high redshifts systems such as the quasar emission or absorption spectra. 
This could proceed along parallel lines, providing alternative 
interpretations to the already developed analysis of the possible 
time dependence of the proton-electron mass ratio \cite{Flambaum, Murphy}, 
acquiring information about the time evolution of the Higgs field.
Based on the alternative assumption that quasars redshifts are not necessarily 
of cosmological origin, the possibility that quasars are naked singularities 
\cite{deFelice,Joshi,Joshi1} with strong gravitational redshifts possibly 
containing also a Higgs component should be left open as a possibility.
Astrophysical limits arising from the analysis as suggested here are 
critical to test proposals that rely on having the Higgs boson being 
responsible for the inflationary model, as discussed in \cite{Bezrukov}, 
especially considering the large curvature-Higgs coupling (order of 
$\xi \simeq 10^4$) required in this scenario.

Furthermore, it is worth noticing that for the conformal coupling 
the vacuum expectation value is increased in a curved spacetime 
corresponding to positive ($R>0$ or $K_1 >0$) scalar invariants. 
Conceptually, the presence of a positive feedback on the Higgs 
expectation value due to a finite curvature may lead to gravitational 
instabilities. If we consider a test mass located near a source 
of curved spacetime, due to the Higgs field its mass will increase 
with respect to the flat spacetime, consequently increasing 
the local curvature, which in turn will increase the value 
of the test mass. In principle, this positive feedback 
mechanism could generate a conceptual issue for the coexistence 
of general relativity and Higgs couplings, at least in its 
nonminimal version. Alternatively, if the feedback turns out 
to be negative, an oscillatory behavior for the spacetime metric 
is expected, leading to a Higgs-driven mechanism for the emission 
of gravitational radiation, with potential implications on the 
spectrum of primordial density fluctuations imprinted in the 
temperature anisotropies of the cosmic microwave background. 
A general analysis on a scalar field with polynomial potential 
terms up to the fourth order has been carried out in \cite{Hosotani}, 
implying $\xi \leq 0$ or $\xi \geq 1/6$ for a stable Higgs field.

Lastly, we want to point out that in the standard model the mass 
of fundamental particles have a different treatment as compared 
to the mass of composite particles. Assuming validity of the 
equivalence principle $-$ an assumption which will be analyzed in 
detail in a future contribution $-$ the gravitational mass of the 
electrons constituting a test body will change if the Higgs field 
is coupled to curvature, while the nucleons will continue to 
keep, at leading order, the usual gravitational charge. 
This is in striking contrast with the standard general relativity 
scenario, whereby all sources of energy contribute without any 
distinctive feature. In turn, this originates an unappealing 
contrast in dealing with the masses, based on their classification 
as fundamental or composite $-$ a classification which has been proven 
to change in time as further layers of elementary particles have emerged.   

\acknowledgments
We are grateful to F. de Felice for fruitful discussions, and H.R. 
Sadeghpour and L. Viola for a critical reading of the manuscript. 
Partial support from the Julian Schwinger Foundation through grant 
JSF 08070000 on Astrophysics of Quantum Vacuum is also acknowledged.

\end{document}